\documentclass[prd,10pt,nofootinbib,tightenlines,showpacs]{revtex4}
\def\journal#1, #2, #3#4, #5#6#7#8    {
    {#1~} {#2}  (#5#6#7#8) #3#4}

\usepackage{amssymb}
\usepackage{amsmath}
\usepackage{lscape}

\begin{document}


\renewcommand{\thesection}{\arabic{section}}
\renewcommand{\thesubsection}{\thesection.\arabic{subsection}}
\renewcommand{\theequation}{\arabic{equation}}
\newcommand{\pv}[1]{{-  \hspace {-4.0mm} #1}}

\baselineskip=12pt 


\begin{center}
{\bf  \Large Geodesic equation in $\kappa$-Minkowski spacetime\\}
 
\bigskip
\bigskip

E. Harikumar {\footnote{e-mail:harisp@uohyd.ernet.in}} \\  
School of Physics, University of Hyderabad, Central University P O, Hyderabad, AP, PIN 500046, India \\[3mm]

T. Juri\'c 
{\footnote{e-mail: tjuric@irb.hr}}, S. Meljanac {\footnote{e-mail: meljanac@irb.hr}},
 \\  
Rudjer Bo\v{s}kovi\'c Institute, Bijeni\v cka  c.54, HR-10002 Zagreb,
Croatia \\[3mm]


\end{center}
\setcounter{page}{1}


{
In this paper, we derive corrections to the geodesic equation due to the $\kappa$ deformation of curved spacetime, up to the first order in the deformation parameter $a$. 
This is done by generalizing the method from our previous paper \cite{elec}, to include curvature effects. We show that the effect of  $\kappa$-noncommutativity can be interpreted as an extra drag that acts on the particle while moving in this $\kappa$-deformed curved space. We have  derived the Newtonian limit of the geodesic equation and using this, we discuss possible bounds on the deformation parameter. We also derive the generalized uncertainty relations valid in the non-relativistic limit of the $\kappa$-spacetime.
}




\section{Introduction}

Noncommutative space entered physics in 1947 when Snyder proposed a model of noncommutative spacetime, admitting a fundamental length \cite{Snyder} as a solution for high energy cutoff, as envisaged by Heisenberg. Since then, motivations for investigating noncommutativity have changed and now noncommutative geometry provides a possible paradigm to capture spacetime uncertainty. Since such uncertainties are encountered in approaches to microscopic theory of gravity\cite{Dopl}, noncommutative geometry naturally comes in the discussions of quantum gravity. It is known that in the low energy limit, the symmetry algebra of certain quantum gravity models is the $\kappa$-Poincare algebra.The corresponding spacetime, known as $\kappa$-Minkowski spacetime is an example for a Lie algebraic type noncommutative spacetime\cite{JL1,JL2}. $\kappa$-Minkowski spacetime has been studied in the context of deformed special relativity also\cite{KG,AC,Ghosh}. Various aspects of this spacetime have been brought out in\cite{md,sm,sm1}. 

In recent times, different field theory models on $\kappa$-spacetime have been constructed, using various approaches, and many interesting aspects of these models have been analyzed\cite{JL3,sm2,jw1}. Particularly, scalar field theory on $\kappa$-Minkowski space is analyzed in \cite{kscal}. By investigating the effect of $\kappa$-deformation parameter on different physical models and comparing their predictions with well known experimental/observational results, bounds on this noncommutative parameter have been obtained by various authors\cite{bol,ab,msk,akk}.

One of the motivations of studying noncommutative geometry is that it naturally encodes the quantum structure of the spacetime. Thus it is of intrinsic interest to see how gravity theories can be constructed on non-commutative spacetime(s) and to analyze how these models differ from the gravity theories in the commutative spacetime. Several authors have studied these issues, using many different approaches\cite{ncgrav1,ncgrav2,ncgrav3, vor,ksg}. Some of these authors have constructed gravity theories on Moyal spacetime, using $*$-product approach. Tetrad formulation of general relativity was generalized to noncommutative case, leading to complex gravity models.  In \cite{ncgrav3}, notions of Hopf algebra was used to construct a noncommutative diffeomorphism invariant theory and different aspects of these models were analyzed. In\cite{vor}, by demanding the noncommutative parameter $\theta^{\mu\nu}$ to be covariantly constant, a generalized $*$ product on the curved noncommutative spacetime was obtained. Using this, a possible generalization of gravity to noncommutative spacetime was studied and modification to geodesic equation was obtained. In \cite{ksg}, behaviour of scalar field near a black hole in $\kappa$-spacetime was investigated.

In this paper, we derive the geodesic equation on the $\kappa$-spacetime, valid up to first order in the deformation parameter. We use a generalization of Feynman's approach\cite{Dyson,Tanimura,minfeyn},  in deriving the geodesic equation in $\kappa$-spacetime. It was shown that the homogeneous Maxwell's equations can be derived by starting with the Newtons force equation and the (assumed) commutators between the coordinates and velocities\cite{Dyson}, which has been generalized to relativistic case in \cite{Tanimura}. In \cite{Tanimura}, it was shown that the consistent interactions possible for a relativistic particle are with scalar, vector and gravitational fields. Various aspects of Feynman's approach have been studied  in \cite{abyg, abyg1}. This method has been generalized to Moyal space time in \cite{ncfey} and to $\kappa$-spacetime in \cite{eh,elec}.

In \cite{elec}, we have generalized the approach of \cite{minfeyn} to $\kappa$-spacetime,  and derived the $\kappa$-deformed Maxwell's equations and Lorentz equation, valid up to first order in the deformation parameter-$a$ and its classical limits were obtained.  We found that the modified Newton's equation depends on  velocities and this effect can be interpreted as due to a background electromagnetic field. In the case of deformed Lorentz equation, we have quadratic terms in velocities (apart from the linear ones). These can be interpreted as due to the curvature induced by the $\kappa$-deformation of the spacetime. Similar feature was shown in the case of Moyal spacetime in\cite{vor1}.  In\cite{elec}, we have found that the electrodynamics depends on the mass of the particle (apart from its charge). We have also investigated the trajectory of the charged particle in a constant electric field in $\kappa$-spacetime, showing the effect of induced electromagnetic field and/or curvature. Thus it is natural to ask what happens if we consider the motion of particle in  $\kappa$-spacetime with curvature. We take up this issue of constructing the geodesic equation in this paper.

This paper is organized as follows. In the next section, we generalize the method of \cite{Tanimura}, along the lines taken in \cite{elec}, to derive the geodesic equation in the commutative spacetime. This approach is suitable for generalization to $\kappa$-spacetime with non-trivial metric. In section 3, we generalize this approach to $\kappa$-deformed spacetime with curvature. Here first, we adapt the method of section 2 to the case of $\kappa$-Minkowski spacetime.  This is discussed in section 3.1. Then in section 3.2, geodesic equation is derived in the $\kappa$-spacetime with curvature. We show that the $\kappa$-dependent correction to geodesic equation is cubic in velocities. Then in section 3.3, we obtain the correction to the Newtonian limit of the geodesic equation. We see that only the radial force equation gets a $\kappa$ dependent modification. Using this modification, we discuss possible bounds on the deformations parameter $a$. In section 3.4, we derive the non-relativistic correction to the commutation relations. Using this, we get the generalized uncertainty principle.  Our concluding remarks are given in section 4.

We work with $\eta_{\mu\nu}=(+,-,-,-)$.


\section{Gravity and Feynman approach}

It was  shown that in the flat spacetime,  Feynman approach and minimal coupling method are equivalent \cite{minfeyn,elec} and that one can derive the general equation of motion for a charged particle. The same was done for $\kappa$-Minkowski space \cite{elec}. In \cite{Tanimura} Feynman approach was generalized for the case of general relativity, resulting in the derivation of the geodesic equation. We are going to revise the approach taken in \cite{Tanimura}, along the lines taken in\cite{elec}, in a way that we could generalize the procedure for $\kappa$-Minkowski space. The goal is to get corrections to the geodesic equation due to the $\kappa$-deformation of spacetime, up to the first order in the deformation parameter $a$. We know that a  relativistic particle of mass $m$ and electric charge $e$ is described by $x_{\mu}(\tau)$ in $4D$-Minkowski space, where $\tau$ is a parameter and that one can write the following  relations
\begin{equation}\begin{split}\label{1}
&[x_{\mu}(\tau),x_{\nu}(\tau)]=0,  \qquad [x_{\mu}(\tau),p_{\nu}(\tau)]=-i\eta_{\mu\nu}, \\
F_{\mu}&=\partial _{\mu}\phi+eF_{\mu\nu}\dot{x}^\nu,  \qquad F_{\mu\nu}=\partial _{\mu}A_{\nu}(x)-\partial _{\nu}A_{\mu}(x),\\
\end{split}\end{equation}
where $p_{\mu}=m\dot{x}_{\mu}+eA_{\mu}$ is the canonical momentum operator, $F_{\mu}=m\ddot{x}_{\mu}$ the force, $F_{\mu\nu}$  the electromagnetic strength tensor, $A_{\mu}$ is a gauge field and $\phi(x)$ is an arbitrary function of $x$.
In this paper, we are concerned with gravity only, and hence we do not need the gauge field. That is to say, we are analyzing particles with no electric charge. So, for a neutral particle we have
\begin{equation}\begin{split}\label{2}
[x_{\mu}&,x_{\nu}]=0,  \qquad [p_{\mu},p_{\nu}]=0, \\
& \   \   \  [x_{\mu},p_{\nu}]=-i\eta_{\mu\nu}\\
&F_{\mu}=0,   \qquad  p_{\mu}=m\dot{x}_{\mu},\\
\end{split}\end{equation}\label{2}
where we have taken $\phi(x)=0$, because this choice will lead to the correct geodesic equation. In \cite{Tanimura} it is argued that the generalization from flat to curved space can be done by the argument that Eq.(\ref{1}) are to be taken as valid in a local Lorenz frame of reference and  effect of gravity can be brought in by replacing the Minkowskian metric $\eta_{\mu\nu}$ with an arbitrary metric $g_{\mu\nu}(X)$. It is showed that this assumption leads to geodesic equation. So we also postulate 
\begin{equation}\label{metric}
[X_{\mu},X_{\nu}]=0 \qquad [X_{\mu},P_{\nu}]=-ig_{\mu\nu}(X),
\end{equation}
but for now we are assuming only that the `metric' $g_{\mu\nu}(X)$ is a function of operator $X$ and that it is a symmetric tensor (which is implied by the first equation in (\ref{metric}) ). All the shifting of indices is done with $\eta_{\mu\nu}$(this is different from what is done in \cite{Tanimura}). $X_{\mu}(\tau)$ is a new position operator and $P_{\mu}(\tau)$ is the corresponding conjugate momenta, $m\dot{X}_{\mu}=P_{\mu}$, and we want to solve (\ref{metric}) in terms of the operators given in (\ref{2}). It is easy to see that we can construct operators $X$ and $P$ as follows 
\begin{equation}\label{ansatz}
X_{\mu}\equiv x_{\mu}, \qquad P_{\mu}\equiv g_{\mu\alpha}p^{\alpha},
\end{equation}
where $x_{\mu}$ and $p_{\nu}$ satisfy (\ref{2}). Now, we take the derivative with respect to $\tau$ of Eqs.(\ref{metric}) and get
\begin{equation}\label{derive}
\frac{1}{m}[P_{\mu},P_{\nu}]+[X_{\mu},\ddot{X}_{\nu}]=-i\frac{d g_{\mu\nu}}{d \tau},
\end{equation}
where we have used $\dot{P}_{\mu}=m\ddot{X}_{\mu}$. Using Eqs.(\ref{ansatz}) and (\ref{2}) we have
\footnote{Notice that in the Feynman's approach or its generalization in \cite{Tanimura}, notion of conjugate momentum is not used. Where as, in \cite{minfeyn} and in \cite{elec}, conjugate momentum is used and it has been shown that the Feynman's approach and this method of using conjugate momentum are equivalent.}
\begin{equation}\label{PP}
[P_{\mu},P_{\nu}]=i(g_{\mu\alpha}\frac{\partial g_{\nu\beta}}{\partial x_{\alpha}}-g_{\nu\alpha}\frac{\partial g_{\mu\beta}}{\partial x_{\alpha}})p^{\beta},
\end{equation}
where we have used $[p_{\mu},f(x,p)]=i\frac{\partial f}{\partial x^\mu}$. Also we have
\begin{equation}\label{Dg}
\dot{g}_{\mu\nu}=\frac{\partial g_{\mu\nu}}{\partial X_{\beta}}\dot{X}_{\beta}=\frac{1}{m}\frac{\partial g_{\mu\nu}}{\partial X_{\beta}}P_{\beta}=\frac{1}{m}\frac{\partial g_{\mu\nu}}{\partial x_{\beta}}g_{\beta\alpha}p^{\alpha}
\end{equation}
Equations (\ref{derive}), (\ref{PP}), (\ref{Dg}) give
\begin{equation}\label{ddX}
[X_{\mu},m\ddot{X}_{\nu}]=-\frac{i}{m}(g_{\mu\alpha}\frac{\partial g_{\nu\beta}}{\partial x_{\alpha}}-g_{\nu\alpha}\frac{\partial g_{\mu\beta}}{\partial x_{\alpha}}+g_{\alpha\beta}\frac{\partial g_{\mu\nu}}{\partial x_{\alpha}})p^{\beta}.
\end{equation}
If the operator $\ddot{X}$ is $\ddot{X}_{\mu}=\ddot{X}_{\mu}(x,p),$ then for the left hand side (L.H.S.) of Eq. (\ref{ddX}) we get
\begin{equation}
[X_{\mu},\ddot{X}_{\nu}]=[x_{\mu},\ddot{X}_{\nu}]=-i\frac{\partial \ddot{X}_{\nu}}{\partial p^{\mu}}.
\end{equation}
Now we can integrate Eq. (\ref{ddX}) over $p^\mu$ and get
\begin{equation}
m\ddot{X}_{\nu}=G_{\nu}+\frac{1}{2m}(g_{\mu\alpha}\frac{\partial g_{\nu\beta}}{\partial x_{\alpha}}-g_{\nu\alpha}\frac{\partial g_{\mu\beta}}{\partial x_{\alpha}}+g_{\alpha\beta}\frac{\partial g_{\mu\nu}}{\partial x_{\alpha}})p^{\beta}p^{\mu}
\end{equation}
where we can choose $G_{\nu}(x)=0$. With
\begin{equation}\label{gamatild}
\tilde{\Gamma}_{\nu\mu\beta}=-\frac{1}{2}(g_{\mu\alpha}\frac{\partial g_{\nu\beta}}{\partial x_{\alpha}}-g_{\nu\alpha}\frac{\partial g_{\mu\beta}}{\partial x_{\alpha}}+g_{\alpha\beta}\frac{\partial g_{\mu\nu}}{\partial x_{\alpha}})
\end{equation}
we finally  get
\begin{equation}\label{tilde}
m\ddot{X}_{\nu}+\frac{1}{m}\tilde{\Gamma}_{\nu\mu\beta}p^{\beta}p^{\mu}=0.
\end{equation}
It is straight forward to see that all the Jacobi identities are satisfied. Eq. (\ref{gamatild}) is similar to the Christoffel symbol in general relativity, and Eq. (\ref{tilde}) is similar to the famous geodesic equation. To make the proper correspondence to gravity we have to go to the classical limit, that is we have to take the limit  $[,]\stackrel{}{\rightarrow}\frac{1}{i}\left\{,\right\}_{PB}$ and  all operators go in to commuting $c$-number functions. We assume that `metric` is invertible and define an inverse of the symmetric tensor $g_{\mu\alpha}$ by the following relation
\begin{equation}\label{inverse}
g_{\mu\alpha}g^{\alpha\nu}=\delta_{\mu} ^\nu .
\end{equation}
From Eqs. (\ref{ansatz}), (\ref{gamatild}) and (\ref{inverse}) we get
\begin{equation}\begin{split}\label{sve}
&g^{\beta\sigma}P_{\sigma}=p^{\beta}, \qquad  g^{\beta\sigma}\frac{\partial g_{\alpha\beta}}{\partial x^{\rho}}=-g_{\alpha\beta}\frac{\partial g^{\beta\sigma}}{\partial x^{\rho}},\\
&\tilde{\Gamma}_{\nu\mu\beta}g^{\beta\sigma}g^{\mu\sigma}=\frac{1}{2}g_{\nu\alpha}(\frac{\partial g^{\alpha\sigma}}{\partial x_{\rho}}+\frac{\partial g^{\alpha\rho}}{\partial x_{\sigma}}-\frac{\partial g^{\rho\sigma}}{\partial x_{\alpha}}) \equiv \Gamma_{\nu} ^{\:\rho\sigma} ,  \\
\end{split}\end{equation}
where $\Gamma_{\nu} ^{\:\rho\sigma}$ is really the Christoffel symbol. Using Eqs. (\ref{sve}) in to (\ref{tilde}) we get the geodesic equation
\begin{equation}\label{geodesic}
\ddot{X}_{\nu}+\Gamma_{\nu} ^{\:\mu\beta}\dot{X}_{\beta}\dot{X}_{\mu}=0.
\end{equation}
Note that all rising and lowering of indices were done using $\eta_{\mu\nu}$, and tensor $g_{\mu\nu}$ is treated only as a symmetric tensor with an inverse defined in Eq. (\ref{inverse}).

Eq. (\ref{geodesic}) illustrates the derivation of the geodesic equation using the  Feynman approach.

\section{$\kappa$-deformation of gravity}
The main results of the paper are discussed in this section. Our aim is to derive the geodesic equation for a particle moving in the noncommutative curved spacetime.  That is, to analyze the $\kappa$ deformations of gravity. With this in mind, we  first discuss the methods of introducing $\kappa$-Minkowski deformations on the flat spacetime. This is then generalized to $\kappa$-deformed spacetime with arbitrary metric. Using the geodesic equation derived, we then study the Newtonian limit and obtain the corrections. Comparing with known observational and experimental results, we discuss the bounds on the deformation parameter suggested by this correction to the Newtonian results. Then we obtain the $\kappa$-modified commutation relations and derive generalized uncertainty relations.

\subsection{$\kappa$-Minkowski space}
$\kappa$-Minkowski space is defined by
\begin{equation}\label{Minkowski}
[\hat{x}_{\mu},\hat{x}_{\nu}]=i(a_{\mu}\hat{x}_{\nu}-a_{\nu}\hat{x}_{\mu}).
\end{equation}
Operators $\hat{x}_{\mu}$ can be realized in terms of operators $x$ and $p$ \cite{elec, kovac} as
\begin{equation}
\hat{x}_{\mu}=x_{\alpha}\varphi^{\alpha} \! _{\mu} (p),
\end{equation}
where $\varphi^{\alpha} \! _{\mu} (p)$ must satisfy
\begin{equation}\label{eqphi}
\frac{\partial \varphi^{\alpha} \! _{\mu}}{\partial p^{\beta}}\varphi^{\beta} \! _{\nu}-\frac{\partial \varphi^{\alpha} \! _{\nu}}{\partial p^{\beta}}\varphi^{\beta} \! _{\mu} =a_{\mu}\varphi^{\alpha} \! _{\nu}-a_{\nu}\varphi^{\alpha} \! _{\mu}
\end{equation}
Solving  Eq. (\ref{eqphi}) up to the first order in deformation parameter $a$ , we get
\begin{equation}\label{phi}
\varphi^{\alpha} \! _{\mu}=\delta^{\alpha} _{\mu}[1+\alpha(a\cdot p)]+\beta a^{\alpha} p_{\mu}+\gamma p^{\alpha} a_{\mu}, \qquad  \alpha,\beta,\gamma \in \mathbb{R}
\end{equation}
where parameters of the realization $\alpha$, $\beta$ and  $\gamma$ have a constraint
\begin{equation}
\gamma-\alpha=1, \qquad \beta \in \mathbb{R}
\end{equation}
It will be convenient for later to define an operator $\hat{y}$ which commutes with $\hat{x}$ (for more on the properties of $\hat{y}$ see \cite{kovac}), i.e.,
\begin{equation}
[\hat{y}_{\mu},\hat{x}_{\nu}]=0 \Longleftrightarrow [\hat{y}_{\mu},\hat{y}_{\nu}]=-i(a_{\mu}\hat{y}_{\nu}-a_{\nu}\hat{y}_{\mu})
\end{equation}
Any function of $\hat{y}$ also commutes with $\hat{x}$
\begin{equation}
[f(\hat{y}),\hat{x}_{\mu}]=0.
\end{equation}
We give $\hat{y}$ and $f(\hat{y})$ up to the first order in $a$
\begin{equation}\begin{split}\label{y}
&\hat{y}_{\mu}=x_{\mu}+\gamma x_{\mu}(a\cdot p)+(\gamma-1)(x\cdot p)a_{\mu}+\beta(x\cdot a)p_{\mu}\\
f(\hat{y})=f(x)&+\gamma(x\cdot \frac{\partial f}{\partial x})(a\cdot p)+(\gamma-1)(a\cdot \frac{\partial f}{\partial x})(x\cdot p)+\beta(a\cdot x)(\frac{\partial f}{\partial x}\cdot p)\\
\end{split}\end{equation}
The canonical  momentum operator (in $e=0$ case) $\hat{p}_{\mu}=m\frac{d \hat{x}_{\mu}}{d \tau}$ is then constructed \cite{elec} as follows
\begin{equation}\begin{split}
&\hat{p}_{\mu}=p_{\alpha}\varphi^{\alpha} \! _{\mu} \longrightarrow \hat{p}_{\mu}=p_{\mu}+(\alpha+\beta)(a\cdot p)p_{\mu}+\gamma a_{\mu}p^{2}\\
&\   \     \       \         \       \      \     \     \     \      \         \      \      \     \    \      \     \    \    \    \    \ [\hat{p}_{\mu},\hat{p}_{\nu}]=0,\\
[\hat{p}_{\mu},\hat{x}_{\nu}]=i&\eta_{\mu\nu}(1+s(a\cdot p)) + i(s+2)a_{\mu}p_{\nu}+i(s+1)a_{\nu}p_{\mu}, \qquad s=2\alpha + \beta.\\
\end{split}\end{equation}
We have shown \cite{elec} that this construction via Feynman approach satisfy all Jacobi identities. The condition that comes by  taking the derivative of Eq. (\ref{Minkowski}), with respect to $\tau$  is
\begin{equation}
[\hat{p}_{\mu},\hat{x}_{\nu}]+[\hat{x}_{\mu},\hat{p}_{\nu}]=i(a_{\mu}\hat{p}_{\nu}-a_{\nu}\hat{p}_{\mu}).
\end{equation}
This completes the results of we need for deriving the geodesic equation in the $\kappa$-Minkowski spacetime.

 \subsection{$\kappa$-dependent corrections to the geodesic equation}

In the flat commutative spacetime, our construction used the conjugate pairs $(x, p)$, and for flat non-commutative spacetime, we had $(\hat{x}, \hat{p})$(see previous section). We showed that all of the operators in the flat non-commutative spacetime could be written in terms of $x$, $p$ and deformation parameter $a$. For the non-commutative spacetime with  curvature,  we will use $(\hat{X}, \hat{P})$ and our main idea is to construct them as functions of $x,~ p$ and deformation parameter $a$. In the case of neutral particles, conjugate momenta is given by  $\hat{P}_{\mu}=m\frac{d \hat{X}_{\mu}}{d \tau}$. After obtaining the realization for $(\hat{X}$ and $\hat{P})$ in terms of $x$ and $p$, the consistent with $\kappa$ generalization of the relationsin Eqs (\ref{metric}) and Eq.( \ref{ansatz}), we derive the corrections to the geodesic equation due to the $\kappa$-deformation of spacetime.

 We start with the postulate
 \begin{equation}
 [\hat{X}_{\mu},\hat{X}_{\nu}]=i(a_{\mu}\hat{X}_{\nu}-a_{\nu}\hat{X}_{\mu}),\label{kappacom}
 \end{equation}
 where 
 \begin{equation}
 \hat{X}_{\mu}=X_{\alpha}\varphi^{\alpha} \! _{\mu},
 \end{equation}
and $\varphi^{\alpha} \! _{\mu}$ satisfies Eq. (\ref{eqphi}). In order to construct $\hat{P}_{\mu}$ we have to satisfy all the Jacobi identities and 
\begin{equation}\label{condition}
[\hat{P}_{\mu},\hat{X}_{\nu}]+[\hat{X}_{\mu},\hat{P}_{\nu}]=i(a_{\mu}\hat{P}_{\nu}-a_{\nu}\hat{P}_{\mu}).
 \end{equation}
We know that in the limit $a\longrightarrow 0,$ we must  have 
\begin{equation}\begin{split}
\hat{X}_{\mu}\longrightarrow &X_{\mu}=x_{\mu} \qquad  \hat{P}_{\mu}\longrightarrow P_{\mu}=g_{\mu\alpha}p^{\alpha}\\
&[\hat{X}_{\mu},\hat{P}_{\nu}]\longrightarrow g_{\mu\nu}(x)\\
\end{split}\end{equation}
In the limit $g_{\mu\nu}(x)\longrightarrow \eta_{\mu\nu}$ we must get
\begin{equation}\begin{split}
\hat{X}_{\mu}\longrightarrow\hat{x}_{\mu} &=x_{\alpha}\varphi^{\alpha} \! _{\mu} \qquad  \hat{P}_{\mu}\longrightarrow \hat{p}_{\mu}=p_{\alpha}\varphi^{\alpha} \! _{\mu}\\
 &[\hat{X}_{\mu},\hat{P}_{\nu}]\longrightarrow [\hat{x}_{\mu},\hat{p}_{\nu}]\\
 \end{split}\end{equation}
Taking these into account, we construct $\hat{P}_{\mu}$. By analogy with Eq. (\ref{metric}) and Eq.( \ref{ansatz}), we just need to  substitute $g_{\mu\nu}(x)$ with a function that commutes with $\hat{x}_{\mu}$, that is with $g_{\mu\nu}(\hat{y})$. Thus we have
\begin{equation}
\hat{P}_{\mu}\equiv g_{\alpha\beta}(\hat{y})p^{\beta}\varphi^{\alpha} \! _{\mu}.
\end{equation}
Using Eq. (\ref{eqphi}), it is straight forward to see that this construction satisfies all Jacobi identities and Eq. (\ref{condition}) is by satisfied to all orders in $a$. For more on construction of $\hat{P}_{\mu}$ see Appendix A. 
 
 So, finally we have
\begin{equation}\label{Def}
\hat{X}=x_{\alpha}\varphi^{\alpha} \! _{\mu} \qquad  \hat{P}_{\mu}=g_{\alpha\beta}(\hat{y})p^{\beta} \varphi^{\alpha} \! _{\mu},
\end{equation}
\begin{equation}\label{XMin}
 [\hat{X}_{\mu},\hat{X}_{\nu}]=i(a_{\mu}\hat{X}_{\nu}-a_{\nu}{X}_{\mu}) ,
\end{equation}
\begin{equation} \label{XP}
[\hat{X}_{\mu},\hat{P}_{\nu}]=-ig_{\alpha\beta}(\hat{y})\left(p^{\beta} \frac{\partial \varphi^{\alpha} \! _{\nu}}{\partial p^{\sigma}}\varphi^{\sigma} \! _{\mu} +\varphi^{\alpha} \! _{\nu} \varphi^{\beta} \! _{\mu} \right)
\end{equation}
Eqs. (\ref{Def}, \ref{XMin}, \ref{XP}) are completely non-perturbative results valid to all orders in $a$.
 
Now we take the derivative of Eq. (\ref{XP}) with respect to $\tau$ , and by using $\frac{d \hat{P}_{\mu}}{d \tau}=m\hat{\ddot{X}}_{\mu}$ and $\frac{d p_{\mu}}{d \tau}=0,$ we get
 \begin{equation}\label{comforc}
[\hat{X}_{\mu},m\hat{\ddot{X}}_{\nu}]=-\frac{1}{m}[\hat{P}_{\mu},\hat{P}_{\nu}]-i\frac{d g_{\alpha\beta}(\hat{y})}{d \tau}\left(p^{\beta} \frac{\partial \varphi^{\alpha} \! _{\nu}}{\partial p^{\sigma}}\varphi^{\sigma} \! _{\mu} +\varphi^{\alpha} \! _{\nu} \varphi^{\beta} \! _{\mu} \right)
 \end{equation}
This is as far as we can get non-perturbativly. We can calculate the right hand side (R.H.S.) of Eq. (\ref{comforc}) explicitly up to the first order in the deformation parameter $a,$ using Eqs. (\ref{phi}, \ref{y}, \ref{Def}). Since our goal is to get the corrections to $\hat{\ddot{X}}_{\mu}$, we assume
 \begin{equation}\label{assum}
 \hat{\ddot{X}}_{\mu}=\ddot{X}_{\mu}+\delta \ddot{X}_{\mu}(a)+O(a^2),
 \end{equation}
where $\delta \ddot{X}(a)$ is linear in $a$ and  generally a function of $x$ and $p$, while $\ddot{X}_{\mu}$ satisfies Eq. (\ref{tilde}). For the L.H.S of Eq. (\ref{comforc}) we have
\begin{equation}\label{com1}
[\hat{X}_{\mu},\hat{\ddot{X}}_{\nu}]=[\hat{X}_{\mu},\ddot{X}_{\nu}]+[X_{\mu},\delta \ddot{X}_{\nu}(a)]+O(a^2).
\end{equation}
 Combining Eq.(\ref{comforc}) and Eq.(\ref{com1}) we have
 \begin{equation}\label{com2}
m[X_{\mu},\delta\ddot{X}_{\nu}(a)]=-[\hat{X}_{\mu},m\ddot{X}_{\nu}]-\frac{1}{m}[\hat{P}_{\mu},\hat{P}_{\nu}]-i\frac{d g_{\alpha\beta}(\hat{y})}{d \tau}\left(p^{\beta} \frac{\partial \varphi^{\alpha} \! _{\nu}}{\partial p^{\sigma}}\varphi^{\sigma} \! _{\mu} +\varphi^{\alpha} \! _{\nu} \varphi^{\beta} \! _{\mu} \right)
 \end{equation}
where the L.H.S. can be writen as
\begin{equation}\label{varX}
[X_{\mu},\delta \ddot{X}_{\nu}(a)]=[x_{\mu},\delta \ddot{X}_{\nu}(a)]=-i\frac{\partial [\delta\ddot{X}_{\nu}(a)]}{\partial p^{\mu}}.
\end{equation} 
First we calculate the R.H.S of Eq. (\ref{com2}) explicitly up to the first order in $a$, using Eqs.(\ref{tilde}, \ref{phi}, \ref{y}, \ref{Def}). Then by using Eq.(\ref{varX}) we can perform the integration of Eq. (\ref{com2}) and obtain $\delta \ddot{X}_{\nu}(a)$, which with Eq. (\ref{assum}) finally gives
\begin{equation}\label{geodef1}
\hat{\ddot{X}}_{\nu}+\frac{1}{m^2}\tilde{\Gamma}_{\nu\mu\beta}p^{\beta} p^{\mu}=\frac{1}{m^2}\tilde{\Sigma}_{\nu\tau\delta\mu}p^{\tau} p^{\delta} p^{\mu},
 \end{equation} 
where $\tilde{\Gamma}_{\nu\mu\beta}$ is given in (\ref{gamatild}) and $\tilde{\Sigma}_{\nu\tau\delta\mu}$ in Appendix B. We see that the main feature of the $a$-dependent corrections is that it is cubic in $p$, and that it depend on the realization, that is on the parameters $\alpha$, $\beta$ and $\gamma$.
 
If we now go to the classical limit (as described in text after Eq. (\ref{geodesic})) by using Eq. (\ref{inverse}) and Eq. (\ref{Def}) we get
\begin{equation}\begin{split}
&p^{\beta} p^{\mu}=g^{\beta\alpha} g^{\mu\sigma} \hat{P}_{\alpha} \hat{P}_{\sigma}+P_{\sigma_{1}} P_{\sigma_{2}} P_{\sigma_{3}} \left\{...a...\right\}^{\beta\mu\sigma_{1}\sigma_{2}\sigma_{3}}+O(a^2)\\
&p^{\tau} p^{\delta}p^{\mu}=g^{\tau\sigma_{1}}g^{\delta\sigma_{2}}g^{\mu\sigma_{3}}P_{\sigma_{1}} P_{\sigma_{2}} P_{\sigma_{3}}+O(a) 
\end{split}\end{equation} 
and with Eq.(\ref{geodef1}) we have
\begin{equation}\begin{split}\label{ppp}
&\tilde{\Gamma}_{\nu\mu\beta}p^{\beta} p^{\mu}=\Gamma_{\nu} ^{\:\alpha\sigma}\hat{P}_{\alpha} \hat{P}_{\sigma}+\tilde{\Gamma}_{\nu\mu\beta}\left\{...a...\right\}^{\beta\mu\sigma_{1}\sigma_{2}\sigma_{3}}P_{\sigma_{1}} P_{\sigma_{2}} P_{\sigma_{3}}+O(a^2)\\
&\tilde{\Sigma}_{\nu\tau\delta\mu}p^{\tau} p^{\delta} p^{\mu}=\tilde{\Sigma}_{\nu\tau\delta\mu}g^{\tau\sigma_{1}}g^{\delta\sigma_{2}}g^{\mu\sigma_{3}}P_{\sigma_{1}} P_{\sigma_{2}} P_{\sigma_{3}}+O(a^2).
\end{split}\end{equation}
Taking into account that $\hat{P}_{\mu}=m\hat{\dot{X}}_{\mu}$ and Eq. (\ref{ppp}) into
(\ref{geodef1}) we get
\begin{equation}\label{kapageo}
\hat{\ddot{X}}_{\nu}+\Gamma_{\nu} ^{\:\alpha\sigma} \hat{\dot{X}}_{\alpha}\hat{\dot{X}}_{\sigma}=m\Sigma_{\nu} ^{\:\sigma_{1}\sigma_{2}\sigma_{3}} \hat{\dot{X}}_{\sigma_{1}}\hat{\dot{X}}_{\sigma_{2}}\hat{\dot{X}}_{\sigma_{3}}+O(a^{2})
 \end{equation} 
where 
\begin{equation}
\Sigma_{\nu} ^{\:\sigma_{1}\sigma_{2}\sigma_{3}}\equiv
-\tilde{\Gamma}_{\nu\mu\beta}\left\{...a...\right\}^{\beta\mu\sigma_{1}\sigma_{2}\sigma_{3}}
+\tilde{\Sigma}_{\nu\tau\delta\mu}g^{\tau\sigma_{1}}g^{\delta\sigma_{2}}g^{\mu\sigma_{3}}
 \end{equation}
and is more explicitly given in Appendix C. Eq. (\ref{kapageo}) presents corrections to the geodesic equation due to the $\kappa$-Minkowski noncommutativity.
 
\subsection{Newtonian limit}
Here we will investigate the ``Newtonian limit'' of the Eq. (\ref{kapageo}).
From now on we will consider a special case of $\kappa$-Minkowski space, that is, we take $a_{\mu}=(a,\vec{0})$. We define the ``Newtonian limit'' by three requirements \cite{caroll}:
\begin{enumerate}
 \item Particles are moving slowly, so we have 
 \begin{equation}
 \frac{d \hat{X}_i}{d \tau}<<\frac{d \hat{X_0}}{d \tau}.
 \end{equation}
\item Gravitational field is weak and can be considered as perturbation about the flat spacetime metric,i.e.,
 \begin{equation}
 g_{\mu\nu}=\eta_{\mu\nu}+h_{\mu\nu}, \qquad \left|h_{\mu\nu}\right|<<1.
 \end{equation}
 \item Gravitational field is static.
\end{enumerate}
From the definition of the inverse metric, $g_{\mu\nu}g^{\nu\sigma}=\delta_{\mu} ^{\sigma}$, we find that to the first order in $h$, $g^{\mu\nu}=\eta^{\mu\nu}-h^{\mu\nu}$. Using these and keeping  only linear terms in $a$ and $h$ , from eq.(\ref{kapageo}),  we get
\begin{equation}\label{0}
\hat{\ddot{X}}_{0}=0
\end{equation}
\begin{equation}\label{i}
\hat{\ddot{X}}_{i}+\frac{1}{2}\frac{\partial h^{00}}{\partial x^{i}}(\hat{\dot{X}}_{0})^2=m\Sigma_{i}^{000}(\dot{X}_{0})^3
\end{equation}
Using Eq.(\ref{0}) we can easily change the derivatives with respect to $\tau$ to derivatives with respect to $\hat{t}$ in eq.(\ref{i}). At first glance it seems that the last term in eq.(\ref{i}) is not re-parametrization invariant, but since we are keeping only linear terms in $a$ and $h$(hence $\Sigma_{i}^{000}$ is already linear in $a$ and $h$) we have $\dot{X}_{0}=\frac{d t }{d \tau}\approx 1+\frac{1}{2}h_{00}\approx 1+O(h)$ so there is no such problem at all. In the commutative case, for Newtons gravitational force we have $F^{i}=-G\frac{mM}{r^3}x^{i}=\frac{1}{2}\frac{\partial h^{00}}{\partial x^{i}}$. By defining $\hat{F}^{i}=m\frac{d^2 \hat{X}^i}{d \hat{t}^2},$ we finally get
\begin{equation}
\hat{F}^{i}=F^{i}(1-\frac{am}{3}C)\label{kappaF},
\end{equation}
where $C=5\alpha+5\beta+12\gamma$. 
Note that the force equation do get $a$ dependent modification, but there is only radial
force (as in the commutative case). But this radial force has a $a$ dependent
correction and this can be compared with the prediction of Pioneer anomaly. Also notice that the $a$ dependent correction depends on the mass of the test particle. This shows that the equivalence principle is violated. The corrections also depend on the choice of realization, that is parameters $\alpha$, $\beta$, and  $\gamma$.

The form of correction to the force equation obtained above in Eq. (\ref{kappaF}) is exactly in the same form as that obtained in \cite{akk} hence we will get the same bounds on $a$ as obtained in \cite{akk}. Thus the Pioneer anomaly sets a bound $|a|\le 10^{-53}m$ and the violation of equivalence principle sets $|a|\le 10^{-55}m$ (for a body of mass 1 kg).

\subsection{Uncertainty relations}
It is well known that from any quantum theory of spacetime one gets a minimal bound on the localization of particles \cite{padm} and that the noncommutativity of spacetime can in fact account for the modification of the Heisenberg uncertainty relations \cite{kempf}. In \cite{mead} it has been argued that even at the Newtonian level there are modification of the uncertainty relations due to gravity. In order to see modifications of the Heisenberg uncertainty principle in our approach, we first see the nonrelativistic limit of Eqs.(\ref{XMin}, \ref{XP})
\begin{equation}\begin{split}
&[x_{i},x_{j}]=0, \\
&[x_{i},p_{j}]=i\hbar(1+ams)\delta_{ij} ,\\
&[x_{0},p_{0}]=-i\hbar(1+3am(s+1)),
\end{split}\end{equation}
where $s=2\alpha+\beta$. Now, using $\Delta A\Delta B\geq\frac{1}{2}\left|\left\langle [A,B]\right\rangle\right|$, we get
\begin{equation}\begin{split}\label{uncertainty}
&\Delta x_{i}\Delta x_{j}\geq0 \\
&\Delta x_{i}\Delta p_{j}\geq\frac{\hbar}{2}(1+ams) \\
&\Delta E\Delta t\geq\frac{\hbar}{2}(1+3am(s+1))
\end{split}\end{equation}

\section{Conclusion and  outlook}
We have analyzed the effect of $\kappa$ deformation on the motion of a particle in curved spacetime. Although, one can find various approaches that handle noncommutative space and gravity (most of them on Moyal space \cite{ncgrav1,ncgrav2,ncgrav3, vor,ksg}), our approach is one of the first attempts that deals  with effect of gravity on $\kappa$-deformed spacetime. In this paper, we have first generalized Feynman approach from \cite{elec} to reproduce results in \cite{Tanimura}. This then enables us to derive the geodesic equation for the $\kappa$-Minkowski spacetime up to the first order in the deformation parameter $a$. The main difference between commutative  and deformed case is that we have an ``extra'' force that is proportional to $\dot{X}^3$.  This term, which is cubic in velocities, can be interpreted as an extra drag that acts on the particle when moving in a  $\kappa$-deformed curved spacetime. This approach allows one to treat this effects as a perturbation to the commutative, curved spacetime. The principal characteristic of our approach is that all the corrections depend on the choice of realization (parameters $\alpha$, $\beta$ and $\gamma$ ) and on the mass of the test particle. Since photon has $m_{\gamma}=0$, there is no change in geodesic equation for light, and also no change in uncertainty relations, which makes it more difficult to set experimental bounds on deformation parameter $a$. Since for certain quantum gravity models the low energy limit is the $\kappa$-Poincare algebra and corresponding spacetime is $\kappa$-Minkowski, our results can be thought of as the effect of quantum gravity. We have derived the  $a$-dependent correction to the Newtonian limit of the geodesic equation. We see that the Newtonian force/potential remains radial, but depends on the mass of the test particle (as well as $a$).  In the ``special relativistic'' limit (obtained by taking $g_{\mu\nu}\rightarrow\eta_{\mu\nu}$), results obtained here, reproduce various deformed special relativity models, since these models differ from ours only in explicit choice of the parameters $\alpha$, $\beta$ and $\gamma$ (that is in the choice of realization.). It is clear that in this limit we have effects that violate Lorentz symmetry and that there is a change in the dispersion relation. This is analyzed in \cite{ab}. It is interesting to note that the bounds on the deformation parameter $a$ obtained here are same as that obtained in \cite{akk}, where a different realization of the coordinates of $\kappa$-Minkowski spacetime was used.

We have shown that the $\kappa$-deformed commutation relations between phase space variables induce  modified uncertainty relations. There have been investigations on the possible modifications of atomic spectra due to  the generalized uncertainty relations and bounds on deformation parameters were obtained \cite{FB}. Thus it is of interest to study the changes in the spectrum of Hydrogen atom due to the $\kappa$-deformed uncertainty relations we have  in Eq.(\ref{uncertainty}). This will be taken up separately.

In the commutative limit $[X_{\mu},P_{\nu}]$ gives rise to the metric $g_{\mu\nu}$ and by analogy, one could interpret $[\hat{X}_{\mu},\hat{P}_{\nu}]$ (or just the symmetric part of it) as giving rise to noncommutative metric $\hat{g}_{\mu\nu}$. Then it would be interesting to construct noncommutative version of Ricci tensor $\hat{R}_{\mu\nu}$ and Ricci scalar $\hat{R}$ in order to get a Lagrangian that would reproduce Eq.(\ref{kapageo}) by action principal. Question of the invariant line element remains unsolved. This problems are of immense  importance and will be reported elsewhere.

\appendix
\section{}
We have a more general construction of the operator $\hat{P}_{\mu}$ up to the first order in the deformation parameter. By differentiating Eq.(\ref{kappacom}) with respect to $\tau$, we get
\begin{equation}\begin{split}
[&\hat{P}_{\mu},\hat{X}_{\nu}]+[\hat{X}_{\mu},\hat{P}_{\nu}]=i(a_{\mu}\hat{P}_{\nu}-a_{\nu}\hat{P}_{\mu}).
\end{split}\end{equation}
This determines the antisymmetric part of $[\hat{P}_{\mu},\hat{X}_{\nu}]$. We can write
\begin{equation}
[\hat{P}_{\mu},\hat{X}_{\nu}]=\hat{S}_{\mu\nu}+\hat{A}_{\mu\nu}
\end{equation}
where $\hat{S}_{\mu\nu}=\hat{S}_{\nu\mu}$ and $\hat{A}_{\mu\nu}=-\hat{A}_{\nu\mu}$. From Eq. (1) and (2) we get 
\begin{equation}
\hat{A}_{\mu\nu}=\frac{i}{2}(a_{\mu}\hat{P}_{\nu}-a_{\nu}\hat{P}_{\mu}).
\end{equation}
In the limit $a\longrightarrow 0$ we must have
\begin{equation}
[\hat{P}_{\mu},\hat{X}_{\nu}]\stackrel{a\rightarrow 0}{\rightarrow} [P_{\mu},X_{\nu}]=ig_{\mu\nu},
\end{equation}
so, up to the first order in the deformation parameter $a$ we have $\hat{S}_{\mu\nu}=ig_{\mu\nu}+\delta S(a)_{\mu\nu}$ or more explicitly
\begin{equation}
\hat{S}_{\mu\nu}=ig_{\mu\nu}+ia_{\alpha}G_{\mu\nu}^{\:\:\alpha\beta}(x)p_{\beta}+O(a^2).
\end{equation}
Here $G_{\mu\nu}^{\ \ \alpha\beta}=G_{\nu\mu}^{\ \ \alpha\beta}$, and then we get
\begin{equation}\label{generalP}
[\hat{P}_{\mu},\hat{X}_{\nu}]=ig_{\mu\nu}+ia_{\alpha}G_{\mu\nu}^{\:\:\alpha\beta}(x)p_{\beta}+\frac{i}{2}(a_{\mu}\hat{P}_{\nu}-a_{\nu}\hat{P}_{\mu})
\end{equation}
We  get constraints on $G_{\mu\nu}^{\:\:\alpha\beta}(x)$ by requiring that the Jacobi identities must be satisfied up to the first order in $a$. From 
\begin{equation}
[[\hat{X}_{\mu},\hat{X}_{\nu}],\hat{P}_{\lambda}]+[[\hat{X}_{\nu},\hat{P}_{\lambda}],\hat{X}_{\mu}]+[[\hat{P}_{\lambda},\hat{X}_{\mu}],\hat{X}_{\nu}]=0.
\end{equation} 
We get
\begin{equation}
\begin{split}
\label{eqG}
&a_{\alpha}(G_{\mu\nu\:\beta} ^{\:\ \ \alpha}-G_{\mu\beta\:\nu} ^{\:\ \ \alpha})=\\
&\alpha a^{\alpha}\bigg(x_{\beta}\frac{\partial g_{\mu\nu}}{\partial x^{\alpha}}- x_{\nu}\frac{\partial g_{\mu\beta}}{\partial x^{\alpha}} \bigg)+\beta x^{\alpha}\bigg(a_{\beta}\frac{\partial g_{\mu\nu}}{\partial x^{\alpha}}-a_{\nu}\frac{\partial g_{\mu\beta}}{\partial x^{\alpha}} \bigg)+\gamma(x\cdot a)\bigg(\frac{\partial g_{\mu\nu}}{\partial x^{\beta}}-\frac{\partial g_{\mu\beta}}{\partial x^{\nu}} \bigg)+\frac{3}{2}(a_{\nu}g_{\mu\beta}-a_{\beta}g_{\nu\mu})
\end{split}
\end{equation}
Now we can construct $G_{\mu\nu}^{\:\:\alpha\beta}(x)$ from $\eta_{\mu\nu}$, $g_{\mu\nu}$ and $\frac{\partial g_{\mu\nu}}{\partial x^{\alpha}}x_{\beta}$, and symbolically we can write this as
\begin{equation}
G_{\mu\nu\alpha\beta}=\sum_{i} A_{i}(g\cdot\eta)_{\mu\nu\alpha\beta}+\sum_{i}B_{i}(\eta\cdot g\cdot\frac{\partial g}{\partial x}x)_{\mu\nu\alpha\beta}, \qquad  A_{i}, B_{i} \in  \mathbb{R} 
\end{equation}
We get constraints on parameters $A_{i}$ and $B_{i}$ from Eq. (\ref{eqG}) and from the limit $g_{\mu\nu}\longrightarrow \eta_{\mu\nu}$. Finally $G_{\mu\nu\alpha\beta}$ is determined by the parameters $\alpha$ and $\beta$ and four more free parameters. Now it is possible to reconstruct operator $\hat{P}_{\mu}$ from Eq. (\ref{generalP}). This procedure is more general but valid up to the first order in the deformation parameter $a$. The construction where $\hat{P}_{\mu}=g_{\alpha\beta}(\hat{y})p^{\beta}\varphi^{\beta} _{\:\mu}$ is the special case of this general procedure, but we have chosen this special case because it is analogous to the undeformed case, and we have proved that it is valid up to all orders in $a$.
\section{}
\begin{equation}
\tilde{\Sigma}_{\nu\tau\delta\mu}={\cal A}_{\nu\tau\delta\mu}+{\cal B}_{\nu\tau\delta\mu}
\end{equation}
where 
\begin{equation}\begin{split}
&{\cal A}_{\nu\tau\delta\mu}=\frac{1}{3} \Bigg\{ -\tilde{\Gamma}_{(\sigma\delta)\alpha}\left[ 2\alpha\delta^{\alpha} _{[\mu}\delta^{\sigma} _{\nu]}a_{\tau} +
2\beta\delta^{\alpha} _{[\mu}\eta_{\nu]\tau}a^{\sigma}+2\gamma\delta^{\sigma} _{\tau}\delta^{\alpha} _{[\mu}a_{\nu]} \right]\\
&+2\delta^{\alpha} _{[\mu}\delta^{\sigma} _{\nu]} \bigg[ \Big(\alpha(a\cdot\frac{\partial g_{\alpha\beta}}{\partial x})(x_{\tau}\bullet)+\beta(a\cdot x)(\frac{\partial g_{\alpha\beta}}{\partial x^{\tau}}\bullet)+\gamma(x\cdot\frac{\partial g_{\alpha\beta}}{\partial x})(a_{\tau}\bullet)\Big)\frac{\partial g_{\sigma\delta}}{\partial x_{\beta}} +\alpha g_{\alpha\beta} \Big( \left\langle \delta^{\beta} _{\tau} \bullet,(a\cdot\frac{\partial g_{\sigma\delta}}{\partial x})\right\rangle\\
&-(\frac{\partial ^{2} g_{\sigma\delta}}{\partial x^{\beta} \partial x}\cdot a)(x_{\tau}\bullet)+\frac{\partial g_{\sigma\delta}}{\partial x^{\beta}}(a_{\tau}\bullet) \Big)   +\beta g_{\alpha\beta}\left(a^{\beta}(\frac{\partial g_{\sigma\delta}}{\partial x^{\tau}}\bullet)+(a\cdot x)(\frac{\partial ^{2} g_{\sigma\delta}}{\partial x^{\tau} \partial x_{\beta}}\bullet) \right)\\ 
&+\gamma\left(g_{\alpha\beta}\frac{\partial ^{2} g_{\sigma\delta}}{\partial x_{\beta} \partial x^{\lambda}}  x^{\lambda}-\tilde{\Gamma}_{(\delta\sigma)\alpha}\right)(a_{\tau}\bullet)
-(\alpha-\gamma)(a\cdot \frac{\partial g_{\sigma\delta}}{\partial x})(x\cdot \frac{\partial g_{\alpha\tau}}{\partial x}) \bigg]\\
&+ 2\delta^{\lambda} _{[\mu} \delta^{\sigma} _{\nu]} \left[ \beta g_{\alpha\beta}a^{\alpha}\left(\delta^{\beta} _{\tau}\bullet\frac{\partial g_{\sigma\delta}}{\partial x^{\lambda}}+\frac{\partial g_{\sigma\delta}}{\partial x_{\beta}}\eta_{\lambda\tau}\bullet \right)+\gamma a_{\lambda}\left(g_{\alpha\tau}\bullet\frac{\partial g_{\sigma\delta}}{\partial x_{\alpha}} -\tilde{\Gamma}_{(\sigma\delta)\tau}\bullet\right)\right]\Bigg\}
\end{split}\end{equation}
and
\begin{equation}\begin{split}
& {\cal B}_{\nu\tau\delta\mu}=\frac{1}{3} \Bigg\{\delta^{\alpha} _{\nu}\delta^{\beta} _{\mu}\bigg[\frac{\partial g_{\alpha\beta}}{\partial x_{\sigma}}\left(\alpha(a\cdot\frac{\partial g_{\sigma\delta}}{\delta x})x_{\tau}+\beta(a\cdot x)(\frac{\partial g_{\sigma\delta}}{\partial x^{\tau}}\bullet)+\gamma(x\cdot \frac{\partial g_{\sigma\delta}}{\partial x})a_{\tau}\right) +\gamma\Big(x^{\rho}\frac{\partial ^{2} g_{\alpha\beta}}{\partial x_{\sigma} \partial x^{\rho}}g_{\sigma\delta}a_{\tau}-\tilde{\Gamma}_{(\alpha\beta)\delta}a_{\tau}\Big) \\
&+\alpha\left(a^{\rho}x_{\delta}\frac{\partial ^{2} g_{\alpha\beta}}{\partial x_{\sigma} \partial x^{\rho}}g_{\sigma\tau}+(a\cdot \frac{\partial g_{\alpha\beta}}{\partial x})g_{\tau\delta}\right)+\beta\left(a^{\rho}g_{\rho\tau}\frac{\partial g_{\alpha\beta}}{\partial x^{\delta}}+(a\cdot x)\frac{\partial ^{2} g_{\alpha\beta}}{\partial x_{\sigma} \partial x^{\delta}}g_{\sigma\tau}\right)\bigg] -(\alpha+\gamma)a_{\mu}\tilde{\Gamma}_{(\nu\delta)\tau}\\
&-\beta a^{\alpha}\bigg(\tilde{\Gamma}_{(\alpha\tau)\delta}\eta_{\mu\nu}+\tilde{\Gamma}_{(\alpha\nu)\delta}\eta_{\mu\tau}+ \tilde{\Gamma}_{(\alpha\mu)\delta}\eta_{\nu\tau}\bigg)-2\gamma\tilde{\Gamma}_{(\mu\tau)\delta}a_{\nu}-2\alpha a_{\tau}\tilde{\Gamma}_{(\nu\mu)\delta} -\alpha\bigg(x_{\mu}(a\cdot\frac{\partial\tilde{\Gamma}_{\nu\delta\tau} }{\partial x})-2\tilde{\Gamma}_{\nu\mu\delta}a_{\tau}\bigg)\\
&-\beta\bigg((a\cdot x) \frac{\partial\tilde{\Gamma}_{\nu\delta\tau} }{\partial x^{\mu}}-2\tilde{\Gamma}_{\nu\delta\tau}a_{\mu}\bigg)-\gamma a_{\mu}\bigg((x\cdot\frac{\partial\tilde{\Gamma}_{\nu\delta\tau} }{\partial x})-2\tilde{\Gamma}_{\nu\delta\tau}\bigg)\Bigg\}.\\
\end{split}\end{equation}
Here operator $\bullet$ stands for the position where operator $p^{\tau}$ is to be placed in Eq. (\ref{geodef1})
\section{}
\begin{equation}
\Sigma_{\nu} ^{\:\sigma_{1}\sigma_{2}\sigma_{3}}\equiv-\tilde{\Gamma}_{\nu\mu\beta}\left\{...a...\right\}^{\beta\mu\sigma_{1}\sigma_{2}\sigma_{3}}+\tilde{\Sigma}_{\nu\tau\delta\mu}g^{\tau\sigma_{1}}g^{\delta\sigma_{2}}g^{\mu\sigma_{3}}
\end{equation}
where
\begin{equation}\begin{split}
&\left\{...a...\right\}^{\beta\mu\sigma_{1}\sigma_{2}\sigma_{3}}=\\
&-2g^{\epsilon\sigma_{1}}g^{\rho\sigma_{2}}g^{\sigma_{3}(\beta}g^{\mu)\kappa}\bigg[ g_{\alpha\epsilon}(\alpha \delta^{\alpha} _{\kappa} a_{\rho}+\beta a^{\alpha} \eta_{\rho\kappa}+\gamma \delta^{\alpha} _{\rho} a_{\kappa})+\alpha\Big(a\cdot \frac{\partial g_{\alpha\epsilon}}{\partial x}\Big)+\beta(a\cdot x)\frac{\partial g_{\alpha\epsilon}}{\partial x^{\rho}}+\gamma\Big(x\cdot \frac{\partial g_{\alpha\epsilon}}{\partial x}a_{\rho}\Big)\bigg]
\end{split}\end{equation}

\noindent{\bf Acknowledgment}\\
 TJ and SM were supported by
the Ministry of Science and Technology of the Republic of Croatia under contract No.
098-0000000-2865.

\end{document}